# The Lorenz ratio as a guide to scattering contributions to Planckian transport

F. Sun[1], S. Mishra[1], U. Stockert[1], R. Daou[2], N. Kikugawa[3], R. S. Perry[4,5], E. Hassinger[1], S. A. Hartnoll[6], A. P. Mackenzie[1,7,*], V. Sunko[1,8,*]

[1] Max Planck Institute for Chemical Physics of Solids, 01187 Dresden, Germany

[2] Laboratoire de Cristallographie et Sciences des Matériaux (CRISMAT), Norman die Université, UMR6508 CNRS, ENSICAEN, UNICAEN, 14000 Caen, France

[3] National Institute for Materials Science, Ibaraki 305-0003, Japan

[4] London Centre for Nanotechnology and Department of Physics and Astronomy, University College London, London WC1E6BT, United Kingdom

[5] ISIS Neutron and Muon Source, Science and Technology Facilities Council, Didcot OX11 0QX, United Kingdom

[6] Department of Applied Mathematics and Theoretical Physics, University of Cambridge, Cambridge CB3 0WA, UK

[7] School of Physics and Astronomy, University of St. Andrews, St. Andrews KY16 9SS, UK

[8] Department of Physics, University of California, Berkeley, California 94720, USA

[*] Emails: Andy.Mackenzie@cpfs.mpg.de; vsunko@berkeley.edu.

**Abstract**

In many physical situations in which many-body assemblies exist at temperature $T$, a characteristic quantum-mechanical time scale of approximately $\hbar/k_B T$ can be identified in both theory and experiment, leading to speculation that it may be the shortest meaningful time in such circumstances. When this behaviour is investigated by probing the scattering rate of strongly interacting electrons in metals, it is clear that in some cases only electron-electron scattering can be its cause, while in others it arises from high-temperature scattering of electrons from quantised lattice vibrations, i.e. phonons. In metallic oxides, which are among the most studied materials, analysis of electrical transport does not satisfactorily identify the relevant scattering mechanism at 'high' temperatures near room temperature. We employ a contactless optical method to measure thermal diffusivity in two Ru-based layered perovskites, $Sr_3Ru_2O_7$ and $Sr_2RuO_4$, and use the measurements to extract the dimensionless Lorenz ratio. By comparing our results to the literature data on both conventional and unconventional metals we show how the analysis of high-temperature thermal transport can both give important insight into dominant scattering mechanisms, and be offered as a stringent test of theories attempting to explain anomalous scattering.

**Introduction**

The challenge of understanding the widespread observation of so-called Planckian scattering [1] continues to generate interest in many fields of physics. In many different physical situations, a scattering rate $\frac{1}{\tau} \cong \frac{k_B T}{\hbar}$ can be deduced from the measurement of the appropriate transport coefficients [2] [3]. To fully account for these experimental findings, plausible mechanisms for both the *T*-linear temperature dependence and its prefactor need to be identified. Although progress has been made, many mysteries and open questions remain, particularly regarding the reason that the prefactor is so often observed to be close to the simple ratio of the two fundamental constants $k_B$ and $\hbar$.

Study of the transport properties of solids provides an important window on the problem. Indeed, the pronounced *T*-linear d.c. electrical resistivity observed near optimal doping in the copper-oxide high temperature superconductors has been a cause of fascination for over thirty years [4]. Although other experiments such as optical conductivity and photoemission are in principle better probes of the scattering rate of electrons, d.c. resistivity can be carried out in such a wide range of physical conditions (for example ranges of temperature, high magnetic fields, hydrostatic and uniaxial pressure) that resistivity studies have been highly influential in establishing the ubiquity of the *T*-linear scattering rate [5] [6] [7] [8] [9]. Their analysis has highlighted one of the central questions in the field: is the observed Planckian scattering fundamentally associated with equilibration, *i.e.,* with inelastic processes, or not? If it is, fascinating links emerge with bounds deduced on the time rate of growth of quantum chaos in thermal systems [10], and with other deductions made from the application of string theory to condensed matter physics [3]. However, a concrete counter-example has been known since the work of Peierls nearly a century ago. In the electron-phonon problem at high temperatures, the scattering of electrons from phonons is quasi-elastic, and the resistivity results from *T*-linear growth of the scattering cross-section, which does not involve internal equilibration of the electron system *per se* [11] [12].

Several theories for the mechanism yielding Planckian scattering rates have been put forward, some purely electronic [13], while others investigate how electron-phonon scattering might be relevant across a much wider range of temperatures and circumstances than had previously been thought [14] [15] [16] [17]. Motivated on the one hand by this theoretical work, and on the other by experiments showing *T*-linear resistivity down to temperatures below 10 mK [8] where any role of phonons seems implausible, our goal was to try and better understand the relative roles of electron-electron and electron-phonon scattering. To do so, we chose to investigate thermal transport, to which both electrons and phonons contribute. Thermal and electrical transport are not independent, and any theory aiming to explain one should aim to simultaneously explain the other, as has been emphasized in *e.g.,* Ref. [18] and [19].

For our main experiment we selected a material, $Sr_3Ru_2O_7$, which shows *T*-linear resistivity over a broad temperature range and for which accurate knowledge of the Fermi surface topography and Fermi velocities exists [5] [20] [21]. Crucially, there is a plausible microscopic theory for *T*-linear resistivity based on electron-electron scattering [22]: in addition to the large Fermi surface sheets dominating transport, there are shallow (~1 meV) pockets which do not significantly contribute to conductivity but do serve as efficient scatterers. Because of their low Fermi energy they are not subject to the Pauli exclusion principle, and the phase space constraints which usually yield $T^2$ resistivity for electron-electron scattering instead give a *T*-linear dependence. We wanted to investigate whether the electron-electron scattering remains relevant at high temperatures, above 100 K, or whether the electrical transport in this temperature range is dominated by phonon scattering. Our conclusion will be that phonon conduction dominates the heat transport over a wide range of temperatures but that only happens because electron-electron scattering is very strong. We further argue that similar logic applies to the thermal transport in cuprates, which is qualitatively and quantitatively similar to that seen in $Sr_3Ru_2O_7$.

**Experiment**

The primary experimental quantity that we set out to measure is the thermal conductivity, which is of particular interest because of strong evidence from comparative studies of insulating and conducting cuprates that it can be decomposed into electron and phonon contributions [23]. Although thermal conductivity is widely explored using conventional techniques with heaters and spatially separated thermometers, radiation losses can cause significant systematic errors above 100 K unless particular care is taken to build bespoke apparatus with careful radiation shielding to ensure that the sample and its environment are always at similar temperatures. While we relied on a standard setup for our low temperature measurements, we adopted a different strategy to acquire our high temperature data. We used an optical technique [18] [24] to measure thermal diffusivity $D$ (using apparatus described in detail in [25]), and independently measured the heat capacity $c$ of the same sample to extract thermal conductivity $\kappa \equiv cD$. Inverse diffusivity and heat capacity data obtained from a $Sr_3Ru_2O_7$ single crystal, whose growth and characterization are described in [26] [21], are shown in Fig. 1. The two techniques for obtaining $\kappa$ are nicely complementary, because at the low temperatures where traditional thermal conductivity measurements are most reliable, the precision of optically determined thermal diffusivity becomes poor. We therefore combined the two data sets to yield the data shown in Fig. 1c for the thermal conductivity of $Sr_3Ru_2O_7$ from 5 K to 300 K, with the resistivity shown in Fig. 1d.

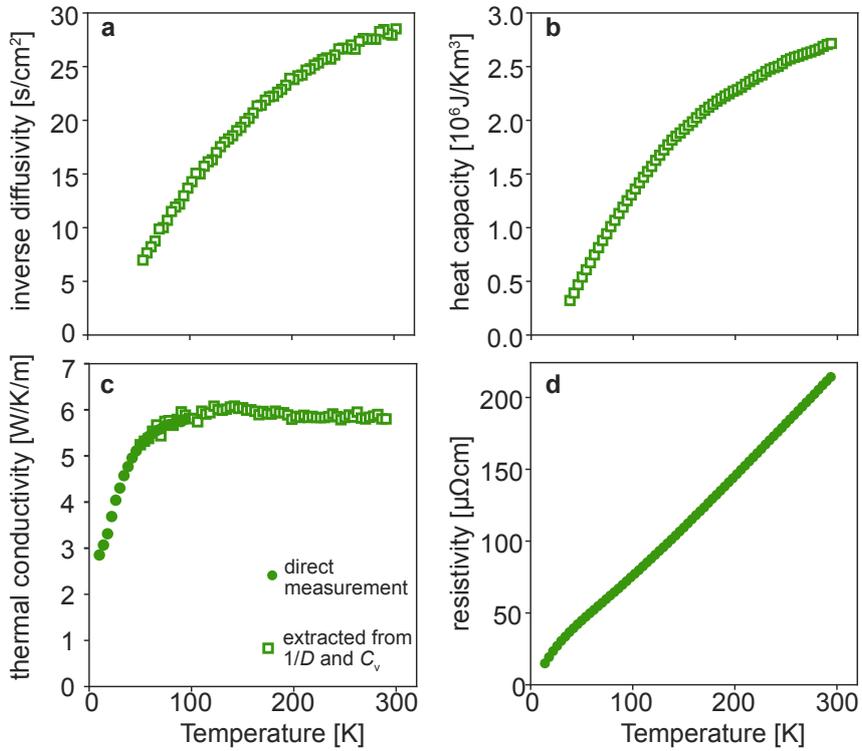

*Fig. 1 a) Inverse thermal diffusivity of $Sr_3Ru_2O_7$ measured using the laser-based technique described in the main text. b) Heat capacity of $Sr_3Ru_2O_7$. c) Thermal conductivity of $Sr_3Ru_2O_7$ measured at low temperatures with a standard one-heater two-thermometer technique (circles) and extracted from the diffusivity and heat capacity data shown in Fig. 1(a,b) (squares). d) Temperature dependent resistivity of $Sr_3Ru_2O_7$, showing the linear behaviour above 75K.*

In the analysis of thermal conductivity of traditional elemental metals, and indeed many strongly correlated metals in the low temperature regime, the Lorenz ratio $L = \kappa\rho/T$ is an often-quoted quantity (for example [27] [28] [29] [30]). Physically, it is a measure of the relative efficiencies of heat and charge transport. In the zero-temperature limit where the heat and electrical currents are both carried by electrons and the elastic impurity scattering dominates, it famously reduces to the universal Lorenz number (sometimes also referred to as the Sommerfeld value) $L_0 = \frac{\pi^2}{3}\left(\frac{k_B}{e}\right)^2 = 2.44 \times 10^{-8}$ V²K⁻². The efficiency of thermal transport is therefore conveniently expressed by the dimensionless quantity $L(T)/L_0$. Although $L(T)/L_0$ can be found plotted up to room temperature for traditional metals in many textbooks (e.g. [31]), it has been plotted over this wide temperature range surprisingly infrequently for correlated electron metals (two examples are Refs [23] and [32]). In Fig. 2 we show it for $Sr_3Ru_2O_7$, along with data extracted from literature thermal conductivity and resistivity of Cu, chosen as a representative elemental metal.

For both metals $\frac{L(T)}{L_0} \cong 1$ at low temperatures, as expected. For Cu, the value of $\frac{L(T)}{L_0}$ returns to $\cong 1$ at room temperature but dips far below 1 at intermediate temperature. This finding can be pictured intuitively as follows. In elemental metals, the carrier concentration is so high, and the Fermi velocity so large, that electrons completely dominate heat transport over the entire temperature range. Away from the impurity-dominated low temperature limit, the main scattering mechanism is electron-phonon scattering. At intermediate temperatures phonons do not have sufficient momentum to relax the electron momentum but are able to degrade their energy, so heat is transmitted less efficiently than charge, causing the Lorenz ratio to drop below one. By 300 K the momentum of a typical phonon is large enough for both the charge and heat currents to be limited by momentum relaxation. The situation resembles that at low temperatures, in the sense that the phonons act like quasi-static scatterers on the characteristic time scale of the electronic motion; in any short-exposure snapshot the material looks like one at low temperatures with a high level of impurity scattering.

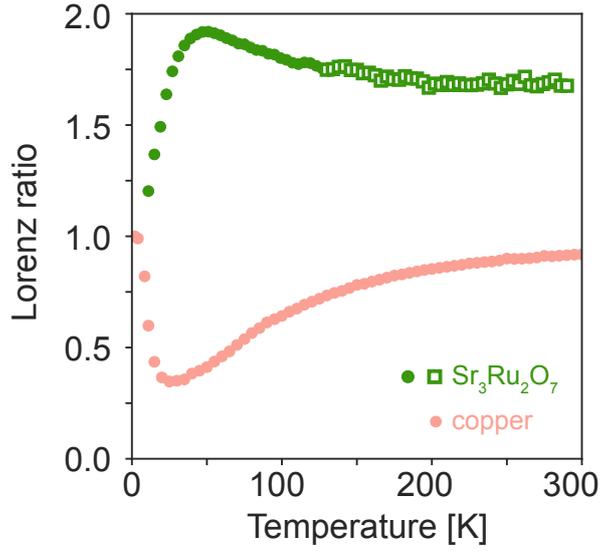

*Fig. 2 Lorenz ratio for $Sr_3Ru_2O_7$ (green), calculated from the data shown in Fig. 1c, and for copper (pink), calculated from Refs. [33] [34]. Circles and squares refer to the Lorenz ratio calculated from the thermal conductivity obtained from the standard one-heater two-thermometer technique and the diffusivity and heat capacity data, respectively.*

The $L(T)/L_0$ data for $Sr_3Ru_2O_7$ differ from those in Cu in two key aspects, one qualitative and one quantitative. At a qualitative level, the dip at intermediate temperatures is replaced by a hump, while from a quantitative point of view $L \to L_0$ at low temperatures but saturates to a considerably higher value above 200 K. Both observations can be rationalized by postulating that, in $Sr_3Ru_2O_7$, phonons as well as electrons make a significant contribution to heat transport, as has been stressed in connection with heat transport in the cuprates [23] [18] [14]. The hump comes because of the rapid growth in the number of excited phonons, while the high temperature saturation reflects the saturation of thermal conductivity, and the enhanced value of $\frac{L}{L_0}$ is most easily understood in terms of an additive phonon contribution to heat transport. Much less discussion has gone into why the phonon contribution is so visible. Here we argue that this observation offers information on the relative contributions of electron-electron and electron-phonon scattering.

A minimal kinetic framework within which to discuss thermal conductivity in a layered metal, like $Sr_3Ru_2O_7$, with a quasi-two-dimensional Fermi surface can be summarized with

$$\kappa = \kappa_{\text{el}} + \kappa_{\text{ph}} = \frac{1}{2} c_{\text{el}} v_{\text{F}}^2 \tau_{\text{el}} + \frac{1}{3} c_{\text{ph}} v_{\text{s}}^2 \tau_{\text{ph}} \qquad (1)$$

where the subscript 'el' refers to electrons, subscript 'ph' to phonons, $c_{el}$ and $c_{ph}$ are the electronic and phononic contributions to the heat capacity, $v_F$ is the average Fermi velocity, $v_s$ is the average sound velocity and the factors of 1/2 and 1/3 appear because the electron (phonon) systems are two- (three-) dimensional, respectively. The expression is based on a quasiparticle picture, but experiments on insulating and metallic cuprates have shown that it is a good starting point for analysis of thermal transport [23] [32]; see also Supplementary Information. Unlike the resistivity, the value and form of $\kappa$, and thus of $\frac{L}{L_0}$, relies not just on the absolute value of $\tau_{el}$ but on the dimensionless ratio $\tau_{el}/\tau_{ph}$. The data indicate that, above 50 K, the phonon contribution to $\kappa$ (and hence to $L$) must be at least as large as the electron one. From inspection of Eq. (1), $\kappa_{ph} \cong \kappa_{el}$ implies that

$$\tau_{ph} \cong \frac{3}{2} \frac{c_{el}}{c_{ph}} \left(\frac{v_F}{v_s}\right)^2 \tau_{el} \qquad (2)$$

All the relevant parameters in Eq. (1) are known with good accuracy in $Sr_3Ru_2O_7$, within the approximations inherent to a quasiparticle analysis of multiband systems [2], allowing a numerical estimate of condition (2). This is done, for $Sr_3Ru_2O_7$ and a number of other materials, in the Supplementary Information. We find that for both ruthenates and cuprates, the scattering rate for electrons must be approximately a factor of 40-50 higher than that for phonons at room temperature to explain the phonon-dominated thermal transport. For Cu, the factor would have to be 2500, which is clearly unattainable and re-emphasizes the fact that its thermal conductivity has to be electron-dominated.

To examine the thermal conductivity contributions quantitatively, it is necessary to consider the way that scattering rates add. Expressions for the scattering rates of electrons and phonons in a system with electron-electron, electron-phonon, phonon-electron and anharmonic phonon-phonon scattering are

$$\frac{1}{\tau_{el}} \cong \frac{1}{\tau_{el-el}} + \frac{1}{\tau_{el-ph}}, \qquad \frac{1}{\tau_{ph}} \cong \frac{1}{\tau_{ph-el}} + \frac{1}{\tau_{ph-ph}} \qquad (3)$$

respectively (with scattering rates adding in both cases because these are separate scattering mechanisms; disorder scattering is neglected since it is temperature independent, and small in these high-purity materials). It is important to note that $\tau_{el-ph}$ and $\tau_{ph-el}$ are not the same; we return to this point later. We have:

$$\frac{\Gamma_{el}}{\Gamma_{ph}} = \frac{\Gamma_{el-el} + \Gamma_{el-ph}}{\Gamma_{ph-el} + \Gamma_{ph-ph}} \geq 50, \tag{4}$$

with $\Gamma = 1/\tau$. Condition (4) could in principle be satisfied in several ways. The first is that the highest scattering rate in the problem is $\Gamma_{el-el}$. The second is that $\Gamma_{el-ph}$ is the dominant scattering rate. $\Gamma_{el-ph}$ is usually larger than $\Gamma_{ph-ph}$ because phonon anharmonicity tends to be weaker than el-ph coupling [35] [36]. $\Gamma_{el-ph}$ is also typically larger than $\Gamma_{ph-el}$ because the phase space for a phonon to decay into a particle-hole pair is smaller than the phase space for an electron to emit a phonon [37]. Although it is a priori unclear if a combination of these effects could allow for a factor of 50 in (4), this seems unlikely: both $\Gamma_{ph-el}$ and $\Gamma_{el-ph}$ are proportional to the electron-phonon coupling constant, so increasing the coupling does not make $\Gamma_{el-ph}$ the dominant scattering rate.

One way to establish if electron-phonon coupling alone could satisfy condition (4) would be to perform a detailed calculation of the relevant scattering cross-sections for realistic Fermi surfaces. To the best of our knowledge such a calculation has not been done yet. For the purposes of this paper, we opted to take a more empirical approach. We identified V$_3$Si as the material with the highest room temperature resistivity (approximately 70 μΩcm [38]) in which resistivity is unambiguously dominated by electron-phonon scattering. Like Sr$_3$Ru$_2$O$_7$ but completely unlike copper, $\Gamma_{el} \geq 50\,\Gamma_{ph}$ if phonons are to dominate the thermal transport (see Supplementary Information), making it at first sight a good candidate for the second scenario. We discovered literature thermal conductivity data taken to 300 K in carefully radiation-shielded apparatus [39]. Combining the thermal conductivity and resistivity to calculate $L(T)/L_0$ yielded the data shown in Fig. 4: in spite of the fact that its room temperature resistivity is a factor of nearly forty higher than that of copper, the Lorenz ratio data of V$_3$Si look qualitatively like those of copper, not those of Sr$_3$Ru$_2$O$_7$. This is a non-trivial observation that strongly suggests that, both qualitatively and quantitatively, a temperature-dependent Lorenz ratio of the kind seen in Sr$_3$Ru$_2$O$_7$ cannot result simply from strong electron-phonon coupling: the 'back-action' simultaneously increases $\Gamma_{ph-el}$, preventing condition (4) from being satisfied. It seems that, in metals, strong electron-electron scattering is necessary to allow for the heat transport to be dominated by phonons.

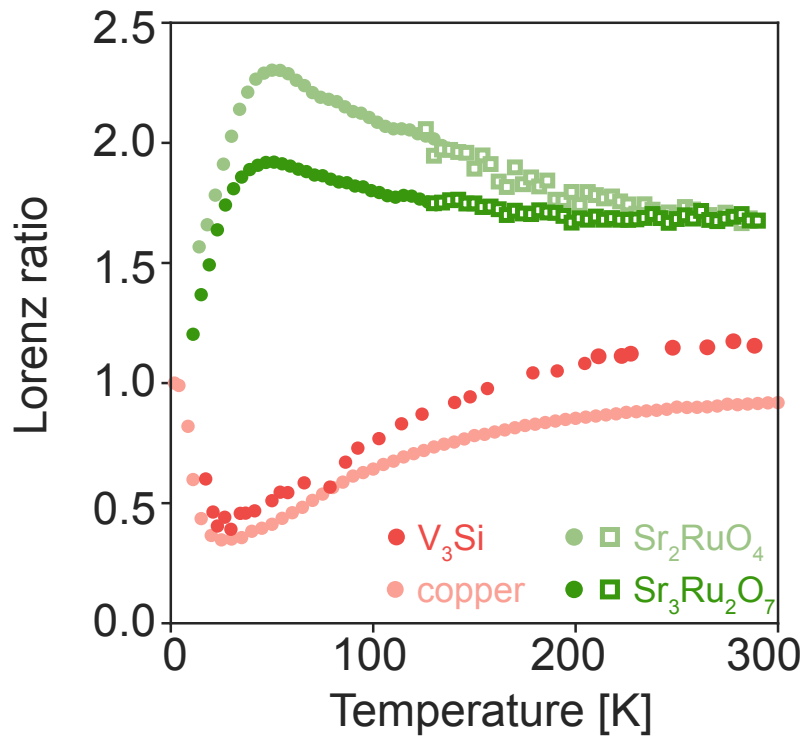

*Fig. 3 Lorenz ratio for $Sr_3Ru_2O_7$ (dark green), $Sr_2RuO_4$ (light green), copper (pink), and $V_3Si$ (red), calculated from Refs. [39] [38]. Square and circles have the same meaning as in Figs. 1 and 2.*

Although our focus in this paper is $Sr_3Ru_2O_7$, we wanted to assess the relevance of our findings to other strongly correlated materials. First of all, in Fig. 3 we show that the Lorenz ratio of the ruthenate $Sr_2RuO_4$, measured using a combination of thermal conductivity, thermal diffusivity and heat capacity measurements, is very similar to that of $Sr_3Ru_2O_7$. Furthermore, we plot the calculated Lorenz ratio data for $La_{2-x}Sr_xCuO_4$ ($x$ = 0.19, 0.20, 0.22) [40] [41] and two samples of $Bi_2Sr_2CaCu_2O_8$ [42] in Fig. 4, again with the results from copper and $V_3Si$. The similarities between cuprates and ruthenates, as well as the difference between those strongly correlated materials and the known electron-phonon scattering dominated materials are very clear.

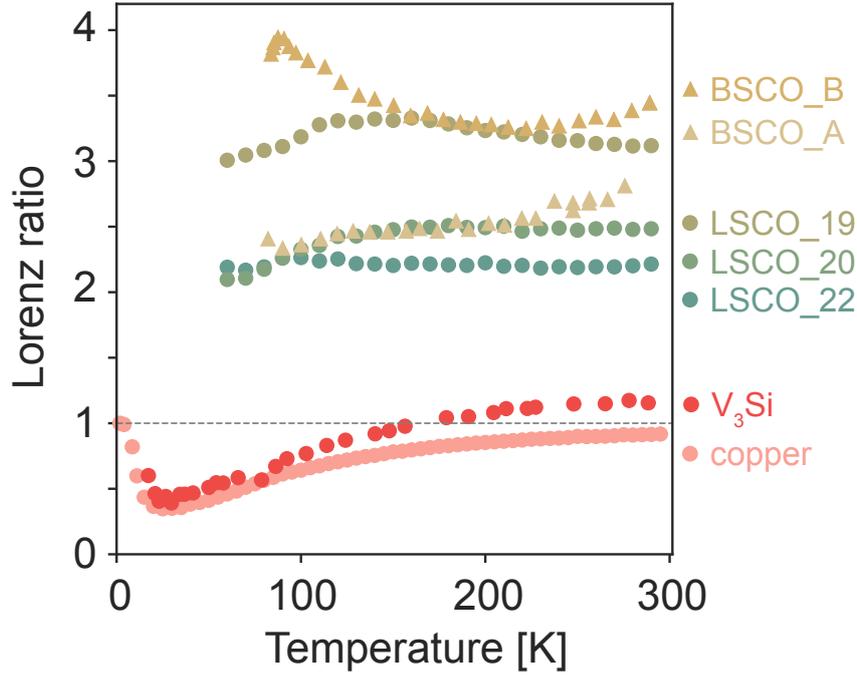

*Fig. 4 Lorenz ratio of $La_{2-x}Sr_xCuO_4$ (x =0.19, 0.20, 0.22) [40] [41] and two samples of $Bi_2Sr_2CaCu_2O_8$ [42], contrasted with that of $V_3Si$ and copper.*

**Discussion**

Figures 3 and 4 summarize our main empirical result. As we have argued above, as long as electrons and phonons are well-defined quasiparticles the most plausible explanation for the key differences between the different material classes is strong electron-electron scattering. It is of course possible that this is not a unique explanation, and that other scenarios such as the strongly coupled electron-phonon soups of the kind recently proposed in Refs. [18] [17], can also account for the data. $L/L_0 > 1$ can also be obtained in some circumstances within interacting electron models that do not explicitly include phonons [43] [44], although other calculations [28] [45], including recent numerical work on the Hubbard model [46] [47] suggested that strong electronic correlations lead to $L/L_0 < 1$, in contrast to the experimental observations. Whatever the final explanation, we hope that we have set a challenge to any theories of Planckian scattering in correlated electron materials such as ruthenates and cuprates: matching the qualitative temperature dependences and magnitudes of the dimensionless quantity $L(T)/L_0$ shown in Figs. 3 and 4 should be added to their goals.

If, as we argue is likely, electron-electron scattering is playing a strong role in producing the observed Planckian resistivity, this leaves open the possibility that equilibration is relevant to the problem. Our analysis neither relies on microscopic details of the electronic scattering [22] [48] [49] nor gives information on the magnitude of the electronic contribution to $L(T)/L_0$ in the higher temperature region in which it is applied. If the underlying electronic contribution is less than one, processes related to electronic equilibration are relevant.

Another fascinating and so far somewhat sparsely studied aspect of the Planckian problem is the way in which different types of scattering combine to create the observed resistivities. We believe that our findings show that things are not as simple as assigning standard electron-phonon scattering as the source of much of the $T$-linear resistivity in $Sr_3Ru_2O_7$ or cuprates: electron-electron processes cannot be ignored. However, that of course does not mean that electron-phonon scattering should be ignored either. It will be interesting to see explicit many-body calculations of the balance between the two such as that recently performed on $Sr_2RuO_4$ [48], and to extend our measurements and analysis to other correlated systems, including strongly overdoped cuprates.

## Methods

**Sample Preparation**

Single crystals of $Sr_3Ru_2O_7$ and $Sr_2RuO_4$ were grown in floating zone furnaces as described in the Refs. [26] and [50].

**Thermal Diffusivity**

Thermal diffusivity was measured with a spatially resolved optical method described in [25] using two laser beams focused on the sample with radii of approximately 2 $\mu$m and separated by a distance $r$ of approximately 20 $\mu$m. The first laser beam is from an Er-doped fiber ultrafast laser with a wavelength of 780 nm and 80 MHz repetition rate which acts as the source of thermal modulation at a frequency of $\omega \sim$ 5 kHz determined by passing it through a mechanical chopper. This causes a periodic and local temperature change, and the heat diffuses radially at a rate which depends on the thermal diffusivity $D$ of the material. The local change in temperature is manifested by a change in temperature dependent reflectivity. The second beam, He-Ne continuous wave laser with a wavelength of 633 nm, probes this change in reflectivity.

A phase lag between the source and the reflected probe beam ($\varphi$) is detected at the frequency ($\omega$) and thermal diffusivity can be calculated from the equation:

$$\varphi = \sqrt{\frac{r^2 \omega}{2D}}$$

This method was used to measure the diffusivity over the broad temperature range of 50 K to 330 K inside a Montana S50 optical cryostat. The sample is mounted on top of a 3-dimensional piezoelectric stage with Attocube nanopositioners which gives the freedom to choose the sample position for measurement and adjustment of the focus.

**Heat Capacity**

The heat capacity was measured using a Physical Property Measuring System (PPMS) from Quantum Design at constant pressure. Samples weighing 8.17 mg ($Sr_3Ru_2O_7$) and 10.23 mg ($Sr_2RuO_4$) were mounted using Apiezon N grease onto a heat capacity puck which has a heater and thermometer, which was then inserted into the cryostat. Measurements were carried out from 10 K to 300 K under high vacuum conditions. The heat capacity is obtained from the relaxation rate of the cooling after the application of a heat pulse to the sample.

**Thermal conductivity in PPMS**

Thermal transport was measured in a PPMS using the Thermal Transport Option (TTO) under high vacuum across the temperature range 2 ~ 300 K, with simultaneous four-point measurement of thermal conductivity and electrical resistivity. Typical dimensions were: spacing between thermal / voltage contacts ~ 1 mm, width 0.5 mm and thickness 0.5 mm. The contacts were made using Dupont 6838 silver paint, cured at 180 °C for about two hours. The agreement of the resistivity measurements with Bruin et al. [5] for $Sr_3Ru_2O_7$ gives evidence that the geometrical uncertainties in our sample mounting and dimension measurement were < 15%. The experimental data from this direct measurement is shown in Supplementary Information Fig. S3.

**Thermal conductivity in bespoke radiation-shielded apparatus**

Thermal conductivity was measured using a standard one-heater, two-thermometer technique in which temperatures were measured using fine wire thermocouples attached to the sample.

The thermal current was measured using a calibrated heat pipe in series with the sample to reduce the error associated with thermal radiation.

**Acknowledgement**

VS is supported by the Miller Institute for Basic Research in Science, University of California, Berkeley. APM and SM acknowledge the support of the Deutsche Forschungsgemeinschaft (DFG) through TRR 288 - 422213477 (project A10). Research in Dresden benefits from the environment provided by the DFG Cluster of Excellence ct.qmat (EXC 2147, project ID 390858940). NK is supported by a KAKENHI Grants-in-Aids for Scientific Research (Grant Nos. 18K04715, 21H01033, and 22K19093) from the Japan Society for the Promotion of Science (JSPS). We thank S.M. Hayden for the useful discussion.

**Supplementary Information**

**S1 *T*-dependence of thermal diffusivity and heat capacity**

**S1.1 Thermal Diffusivity**

As described in the Methods section, the thermal diffusivity was measured with a spatially resolved optical method [1]. Here we show the temperature-dependent inverse diffusivity for both $Sr_3Ru_2O_7$ and $Sr_2RuO_4$ in Fig. S1, respectively. Note that below 50 K, $\varphi$ becomes very small and the signal-to-noise ratio falls below useful levels, so data from that temperature range were not used.

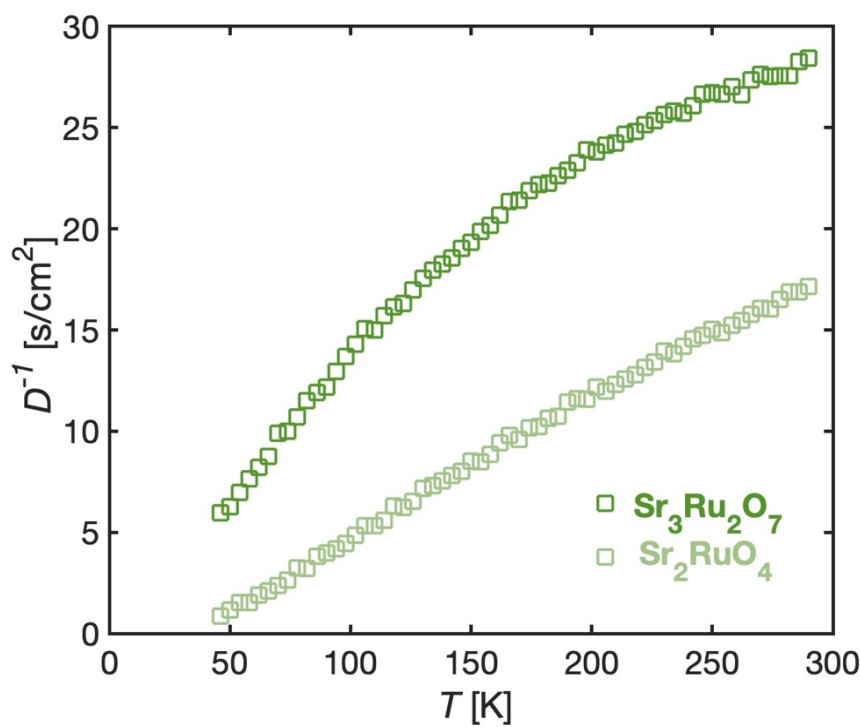

*Fig. S1: $D^{-1}$ of $Sr_3Ru_2O_7$ and $Sr_2RuO_4$ as a function of temperature measured with the optical setup.*

**S1.2 Heat Capacity**

The heat capacity measurements were carried out from 10 K to 300 K under high vacuum conditions. The value of heat capacity is obtained from the relaxation rate of the cooling after the application of a heat pulse to the sample. In Fig. S2, we show the heat capacity as a function of temperature for both $Sr_3Ru_2O_7$ and $Sr_2RuO_4$, respectively.

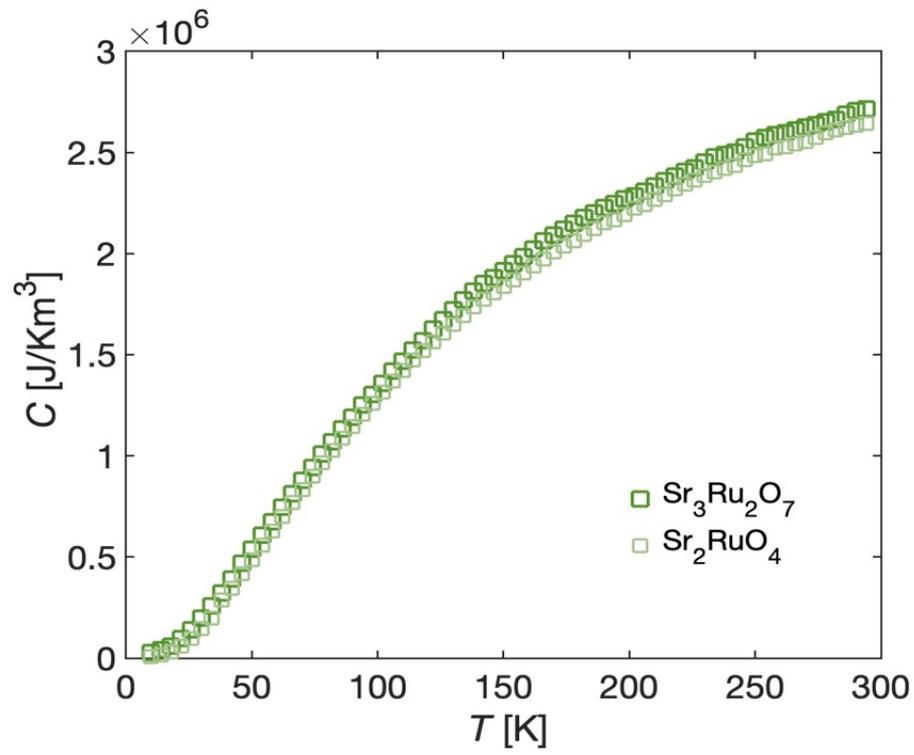

*Fig. S2: Heat capacity of $Sr_3Ru_2O_7$ and $Sr_2RuO_4$ as a function of temperature measured with a PPMS.*

## S2 Comparison between two methods of measuring thermal conductivity

As discussed in the main text traditional thermal conductivity is more reliable at low temperatures below 100 ~ 120 K and becomes challenging at higher temperatures due to radiation losses. In contrast, the uncertainty in the optical measurement of $D^{-1}$ is more at temperatures typically below 50 K as the phase lag is the smallest here and more susceptible to the presence of experimental offsets. Both the techniques have an overlapping temperature regime where both the values are consistent. $\kappa$ is calculated from the thermal diffusivity and heat capacity as $\kappa = cD$, where $c$ is volumetric heat capacity. The measured heat capacity $c_m$ is usually in the units of J/mol Ru K and can be converted to $C$ through the simple conversion:

$$c \text{ (J/K m}^{-3}\text{)} = c_m \text{(J/mol Ru K)} \frac{Z}{N_A V}$$

where, $Z$ is the number of atoms in formula unit and $N_A$ is the Avogadro's constant and $V$ is the volume of the unit cell. Both traditional thermal conductivity measurements and the optical methods are subject to errors of tens of per cent in the absolute values that they yield, so in order to match the data from the PPMS and optical measurements a scaling factor is necessary. We use a scaling factor of 0.59 for $Sr_3Ru_2O_7$ and 0.65 for $Sr_2RuO_4$ for the calculated thermal conductivity to obtain the best match in the region around 100 K where both methods are expected to be subject to small systematic errors. This is a single factor, with no attempt made to fit the shape. In $Sr_3Ru_2O_7$ the good match of the temperature dependence between 50 and 150 K gives confidence in the validity of the use of the scaling factor.

In $Sr_2RuO_4$ the extremely low impurity concentrations lead to very high and somewhat sample-dependent electrical and thermal conductivities below 100 K, and the signal for the optical method is very small. We therefore checked the validity of our high temperature data by direct comparison to the results of the radiation-shielded thermal conductivity performed on a third sample. That crystal was substantially less pure, so the low temperature thermal conductivity shows a much-weakened rise, as expected. However, the optical results agree well with the data from this third sample in the range 100 ~ 300 K that is of primary interest in this paper. (When comparing the two it should be noted that this third sample has its own geometrical uncertainties, and that *no* scale factor was applied to match its data with those from the optical measurements.). The comparison also gives a demonstration of the effects of high temperature radiation losses on PPMS thermal conductivity results. The rise in the data for both $Sr_3Ru_2O_7$ and $Sr_2RuO_4$ above approximately 200 K is the result of this source of systematic error.

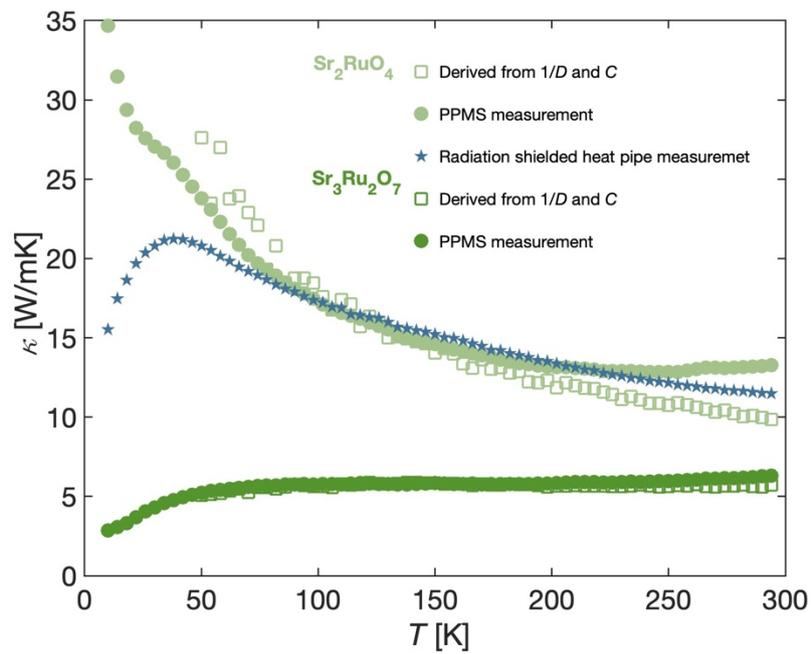

*Fig. S3: Thermal conductivity across the full range of temperature from direct measurement and from optical measurement of thermal diffusivity for $Sr_3Ru_2O_7$ and $Sr_2RuO_4$.*

## S3 Calculation of the Lorentz ratio for V₃Si from literature data

Thermal conductivity ($\kappa$) data were taken from Gladun et al. in Ref. [2], who performed a detailed study of $\kappa$ vs. $T$ for a high quality V₃Si crystal with residual resistivity $\rho_{res}$ = 1.1 µΩcm, using specialized equipment incorporating radiation shielding to ensure accurate data above 150 K.

Resistivity ($\rho$) data were taken from the undoped sample from the work of Caton and Vishwanathan (Ref. [3]), with an interpolation used to smoothly bridge some small gaps in their data, as shown in Fig. S5. Since the sample studied by Caton and Vishwanathan had $\rho_{res}$ = 4.1 µΩcm, we subtracted the temperature-independent value of 3.0 µΩcm from all interpolated data before combining them with the Gladun et al. data to calculate $L = \kappa\rho/T$.

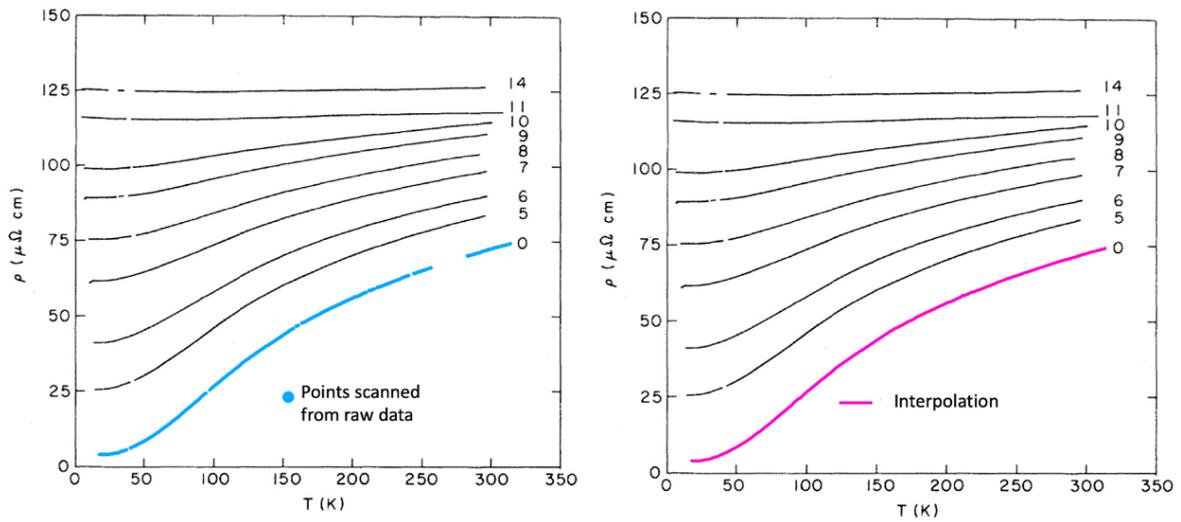

*Fig. S4 Scanned data points (blue) and interpolation function (pink) showing the method used to extract resistivity data from Ref. [3].*

# S4 Comparison of electron and phonon parameters of ruthenates, cuprates and known electron-phonon scattering metals

In this section we give details of the comparisons discussed in the main text between ruthenates, cuprates and two materials, Cu and $V_3Si$, in which electron-phonon processes are thought to dominate the electronic scattering. As outlined in the main text, the quantity, $\frac{3}{d}\frac{c_{el}}{c_{ph}}\left(\frac{v_F}{v_s}\right)^2$, is a measure of the factor by which the total phonon scattering rate must be lower than the total electron scattering rate for phonon and electron contributions to the thermal conductivity to be equal in magnitude. We comment on each column in turn. The parameter $d$ refers to the dimensionality of the electronic system in each material; for every material we assume that the phonon spectrum is best described as being three-dimensional. All data in the remaining columns are at 300 K. To calculate $c_{el}/c_{ph}$ we obtain the electronic specific heat coefficient $\gamma$, either from the literature or our own measurements, and multiply it by 300 to estimate the electronic part $c_{el}$ of the total specific heat $c_{tot}$, which we again obtain either from the literature or our own measurements. The phonon specific heat is then calculated as $c_{ph} = c_{tot} - c_{el}$. Fermi velocity estimates are obtained from either angle-resolved photoemission data or analysis of de Haas – van Alphen effect data, and assumed to be applicable at room temperature. Sound velocities are obtained from the literature. All sources are cited.

| Material | $d$ | $c_{el}/c_{ph}$ | $v_F$ (ms$^{-1}$) | $v_s$ (ms$^{-1}$) | $\frac{3}{d}\frac{c_{el}}{c_{ph}}\left(\frac{v_F}{v_s}\right)^2$ |
|---|---|---|---|---|---|
| $Sr_3Ru_2O_7$ | 2 | 0.36 [4]** | 4.0×10$^4$ [5]† | 4.7×10$^3$ [6] | 40 |
| $Sr_2RuO_4$ | 2 | 0.09 [7]** | 8.5×10$^4$ [8]† | 4.7×10$^3$ [6] | 44 |
| $La_{1.8}Sr_{0.2}CuO_4$ | 2 | 0.01 [9] [10] | 2.3×10$^5$ [10] | 5.9×10$^3$ [11] | 23 |
| $Bi_2Sr_2CaCu_2O_8$ | 2 | 0.01 [12] [13] | 2.5×10$^5$ [13] | 4.3×10$^3$ [14] | 51 |
| Cu | 3 | 0.01 [15] | 1.6×10$^6$ [16] | 3.2×10$^3$ [17] | 2500 |
| $V_3Si$ | 3 | 0.22 [18]** | 1.0×10$^5$ [19]† | 7.0×10$^3$ [20] | 45 |

*Table 1: Experimental values of parameters needed to obtain the ratio between the electronic and phonon scattering rates for different metals. ** Refers to the cases where the value of $c_{tot}$(300 K) was taken from our own measurements and $\gamma$ is used from the cited reference. † Denotes the cases where the values of $v_F$ are deduced from the cited de Haas – van Alphen effect results.*

It is seen that the criterion for observing a significant phonon contribution to the thermal conductivity is so extreme in Cu (chosen as a representative of standard metals) that it will never be reached. This explains why, for those standard metals, both the thermal and electrical conductivity are dominated by electron transport. This need not be the case for the other materials in the table, as long as the total electron scattering rate can be made significantly stronger than the total phonon rate. The most significant finding, as discussed in the main text, is that the criterion for $V_3Si$ is essentially the same as for the strongly correlated cuprates and ruthenates. However, the actual thermal conductivity data for $V_3Si$ are qualitatively more similar to those of Cu than of the strongly correlated materials, suggesting that in $V_3Si$, the 'back-action' of strong phonon-electron scattering prevents the condition $\frac{c_{el}}{c_{ph}}\left(\frac{v_F}{v_s}\right)^2 = 45$ from being satisfied. This in turn suggests that the reason for the observed thermal conductivity of the ruthenates and cuprates is a large electron scattering rate due to a mechanism that gives no back action to the phonon scattering rate. In a naive picture in which electron-electron and electron-phonon scattering can be separated, this suggests that electron-electron processes dominate in the electron scattering in the cuprates and ruthenates, even at room temperature.

## S5 Comments on the quasiparticle-based decomposition of electron and phonon contributions to thermal conductivity

The analysis of thermal conductivity in the main text of this paper is based on equation (1):

$$\kappa = \kappa_{\text{el}} + \kappa_{\text{ph}} = \frac{1}{2} c_{\text{el}} v_F^2 \tau_{\text{el}} + \frac{1}{3} c_{\text{ph}} v_s^2 \tau_{\text{ph}} \qquad (1)$$

As stated in the text, use of this minimal kinetic expression implies the existence of quasiparticles, an assumption that invites scrutiny. However, we believe that past analysis of cuprate data shows that it is in fact a reasonable starting point, for reasons we now explain.

The value of the Lorenz number $L_0$ is straightforwardly derived within the quasiparticle picture:

$$c_{\text{el}} = \frac{\pi}{3} k_B^2 T \frac{k_F}{\hbar v_F} \qquad (2)$$

Inserting in the expression for $\kappa_{\text{el}}$ gives

$$\kappa_{\text{el}} = \frac{\pi^2}{3} k_B^2 T \frac{k_F}{h} v_F \tau_{\text{el}} \qquad (3)$$

while resistivity

$$\rho_{\text{el}} = \frac{h}{e^2 k_F v_F \tau_{\text{el}}} \qquad (4)$$

When the scattering rates for thermal and charge currents are the same, the Lorenz ratio

$$\frac{\kappa_{\text{el}} \rho_{\text{el}}}{T} = \frac{\pi^2}{3} \left(\frac{k_B}{e}\right)^2 = L_0 \qquad (5)$$

While the combination of fundamental constants would be obtained within any approach, the prefactor is a consequence of the quasiparticle analysis.

The cuprates gave the opportunity to check both the separability of the thermal conductivity into electron and phonon contributions and the validity of the value of $L_0$ in experiments combining measurements on insulators with those on their doped, metallic counterparts [21] [22]. In the metals, thermal conductivity was measured directly and compared with the sum of that from the insulator (containing only a phonon contribution) and one calculated from the resistivity using $\kappa_{\text{el}} = L_0 T/\rho_{\text{el}}$. Agreement between the two approaches was good (within approximately 20%).

This is our justification for the use of Eq. (1) as the starting framework for the analysis in the main paper. To the level of accuracy we require, it shows that the quasiparticle-based expressions and the separation of electronic and phononic terms in Eq. (1) give sensible answers in relation to the underdoped cuprates, and there is no reason to suspect that the ruthenates and $V_3Si$ are less conventional.

**References of Supplementary Information**